Scientific Research Publishing

# A Comparison between Abraham and Minkowski Momenta

**Massimo Testa**

Dipartimento di Fisica, Università degli Studi di Roma "La Sapienza", INFN—Sezione di Roma I, Piazzale A. Moro 2, Roma, Italy
Email: massimo.testa@roma1.infn.it

## Abstract

**In this paper I compare the Abraham and the Minkowski forms for the momentum pertaining to an electromagnetic wave inside a dielectric or a magnetic material. The discussion is based on a careful treatment of the surface charges and currents and of the forces acting on them. While in the dielectric case the Abraham momentum is certainly more appealing from the physical point of view, for a magnetic material it suggests an interpretation in terms of magnetic charges and related magnetic currents. The Minkowski momentum for magnetic non conducting materials, on the contrary, has a natural interpretation in terms of an amperian model, in which the dynamics is determined by the Lorentz force acting on bulk and surface electric currents.**

## 1. The Problem

At the beginning of the twentieth century considerable attention has been devoted to the determination of the form of the mechanical momentum to be assigned to an electromagnetic (e.m.) wave propagating inside a given material. Two contrasting solutions to this problem were proposed at the time, the Minkowski espression [1]

$$\boldsymbol{p}_M = \frac{1}{c}\int_{R_\infty} \boldsymbol{D}\times\boldsymbol{B}\,\mathrm{d}\boldsymbol{r} \tag{1}$$

and the Abraham one [2]

$$\boldsymbol{p}_A = \frac{1}{c}\int_{R_\infty} \boldsymbol{E}\times\boldsymbol{H}\,\mathrm{d}\boldsymbol{r}, \tag{2}$$

where $\mathcal{R}_\infty$ denotes the entire three dimensional space. A long time has elapsed since then, but the problem is still under examination in the literature[1] with arguments alternatively in favour of one or the other proposal. The reason for this variety of conclusions is due to the fact that this is a question that finally can only be settled empirically, but experiments are difficult and often do not match the material's idealizations needed to derive the theoretical results.

In Ref. [8] I gave a modest contribution to this debate considering a uniform dielectric $\mathcal{D}$, not necessarily of infinite extension and I explored the consequences of the *assumption*, which looks quite natural, that the forces acting on the polarization charges and currents are to be computed along the same lines as those acting on the "free" ones[2]. Using an expansion in powers of the susceptibility $\alpha$ [10], I showed that this hypothesis favours the Abraham form of the momentum, Equation (2). In particular I examined the so called "Einstein box" argument [11] and the case of oblique incidence on a flat dielectric surface and I was able to show, in particular, that the Snell's law results from the form postulated for the forces, together with the Abraham expression of the e.m. momentum inside the dielectric.

It is the purpose of the present paper to use and extend the methodology presented in [8] in order to compare the different expressions of the forces accompanying the various proposals of the e.m. momentum.

In Section 2 we make some general considerations about the interaction of an e.m. wave with a material body.

In Sections 3, 4 and 5 we discuss the case of a dielectric material, both in the Abraham and the Minkowski setup.

Finally, in Section 6, we extend the analysis to the case of magnetic non conducting materials.

## 2. The Equilibrium of Matter in an e.m. Field

We consider an extended, in general deformable, body, $\mathcal{B}$, surrounded by the vacuum, in the presence of an e.m. wave. $\mathcal{B}$ is subject to e.m. volume forces with a density $\boldsymbol{f}^{(e.m.)}(\boldsymbol{r},t)$ and e.m. surface forces with a surface density $\boldsymbol{f}_\Sigma^{(e.m.)}(\boldsymbol{r},t)$. We also imagine to apply to the body a system of non e.m. external volume forces with a density $\boldsymbol{f}^{(ext)}(\boldsymbol{r},t)$ and external surface forces with a surface density $\boldsymbol{f}_\Sigma^{(ext)}(\boldsymbol{r},t)$.

The equations of motion for this system, together with the action-reaction law, imply

$$\dot{\boldsymbol{p}}^{(e.m.)} + \dot{\boldsymbol{p}}^{(body)} = \boldsymbol{F}^{(ext)}(t), \tag{3}$$

where $\dot{\boldsymbol{p}}^{(e.m.)}$ and $\dot{\boldsymbol{p}}^{(body)}$ are the momenta of the e.m. field and of the body, respectively, and $\boldsymbol{F}^{(ext)}(t)$ is the resultant of the external (non e.m.) forces

$$\boldsymbol{F}^{(ext)}(t) = \int_{\mathcal{B}} \boldsymbol{f}^{(ext)}(\boldsymbol{r},t)\,\mathrm{d}\boldsymbol{r} + \int_\Sigma \boldsymbol{f}_\Sigma^{(ext)}(\boldsymbol{r},t)\,\mathrm{d}\Sigma. \tag{4}$$

If we choose the external forces so that every part of $\mathcal{B}$ is at rest, the body equilibrium conditions

$$f_i^{(e.m.)}(\boldsymbol{r},t) + f_i^{(ext)}(\boldsymbol{r},t) = \partial_k \tau_{ki}(\boldsymbol{r},t) \qquad \boldsymbol{r} \in \mathcal{B} \tag{5}$$

$$f_{\Sigma i}^{(e.m.)}(\boldsymbol{r},t) + f_{\Sigma i}^{(ext)}(\boldsymbol{r},t) = -(n_{\mathcal{B}})_k\, \tau_{ki}(\boldsymbol{r},t) \qquad \boldsymbol{r} \in \Sigma \tag{6}$$

must be satisfied. In Equations (5) and (6) $\tau$ denotes the symmetric matter stress tensor and $\boldsymbol{n}_{\mathcal{B}}$ is the external normal to the body surface.

Integrating Equation (5) over the body volume gives

$$\int_{\mathcal{B}} \left( f_i^{(e.m.)}(\boldsymbol{r},t) + f_i^{(ext)}(\boldsymbol{r},t) \right) \mathrm{d}\boldsymbol{r} = \int_{\mathcal{B}} \partial_k \tau_{ki}(\boldsymbol{r},t)\,\mathrm{d}\boldsymbol{r} = \int_\Sigma (n_{\mathcal{B}})_k\, \tau_{ki}(\boldsymbol{r},t)\,\mathrm{d}\Sigma, \tag{7}$$

while, integrating Equation (6) over the body surface, gives

$$\int_\Sigma \left( f_{\Sigma i}^{(e.m.)}(\boldsymbol{r},t) + f_{\Sigma i}^{(ext)}(\boldsymbol{r},t) \right) \mathrm{d}\Sigma = -\int_\Sigma (n_{\mathcal{B}})_k\, \tau_{ki}(\boldsymbol{r},t)\,\mathrm{d}\Sigma, \tag{8}$$

so that we have

$$\boldsymbol{F}^{(ext)}(t) + \boldsymbol{F}^{(e.m.)}(t) = 0, \tag{9}$$

where $\boldsymbol{F}^{(e.m.)}(t)$ denotes the resultant, volume plus surface, e.m. force acting on $\mathcal{B}$, while $\mathcal{B}$ is kept at rest

[1]For a review of the subject see [3]-[7].

[2]For a recent interesting discussion on this subject see [9].

by the external forces.

Since the material body is in mechanical equilibrium, we have $\boldsymbol{p}^{(body)}=0$ and Equation (3) becomes

$$\dot{\boldsymbol{p}}^{(e.m.)}=\boldsymbol{F}^{(ext)}(t)=-\boldsymbol{F}^{(e.m.)}(t). \tag{10}$$

Given an expression for $\boldsymbol{p}^{(e.m.)}$, Equation (10) allows us to compute the resultant of the forces to be applied to $\mathcal{B}$, in order that the material body stays motionless.

$\boldsymbol{F}^{(ext)}(t)$ is a very natural physical quantity, ideally measured by a static dynamometer, but, for conventional reasons, we will find it convenient to consider $\boldsymbol{F}^{(e.m.)}(t)$, instead.

It is important to notice that, following the procedure just described, we do not need to restrict in any way the entity of the body stresses, provided we use, in the formulas we get, the dielectric and magnetic polarizabilities pertaining to the stressed body in equilibrium. In the following, in order not to make the presentation too heavy, we will consider the simplified situation in which the stressed deformed body is uniform with respect to its electric and magnetic properties. The most general, non uniform case, can be easily treated along the same lines.

In the following sections we will examine the consequences of assuming for $\boldsymbol{p}^{(e.m.)}$ the Abraham or Minkowski form, Equations (1) and (2), both in the case of dielectric and magnetic, non conductive materials.

Although it would be possible to schematize the transition of dielectric or magnetic polarizabilities from the material body to the vacuum in a continuous way, we will find it convenient to consider the electric and magnetic properties of the material body as homogeneous, with a discontinuity at the body surface.

## 3. Dielectric Materials

A uniform linear[3] dielectric $\mathcal{D}$, not necessarily of infinite extension, is a material in which a polarization field is present

$$\boldsymbol{P}(\boldsymbol{r},t)=(\epsilon-1)\boldsymbol{E}(\boldsymbol{r},t)\equiv\alpha\boldsymbol{E}(\boldsymbol{r},t), \tag{11}$$

which defines the (constant) electric susceptibility $\alpha$.

Given $\boldsymbol{P}$, in general we find in $\mathcal{D}$

- polarization charges, with a bulk volume density

$$\rho_p(\boldsymbol{r},t)=-\nabla\cdot\boldsymbol{P}(\boldsymbol{r},t)=-\alpha\nabla\cdot\boldsymbol{E}(\boldsymbol{r},t); \tag{12}$$

for simplicity we consider the case in which the uniform dielectric does not contain free charges[4]. In this case we have $\rho_p(\boldsymbol{r},t)=0$;

- surface charges with a surface density [13] [14] on its boundary $\Sigma$,

$$\tilde{\sigma}_p(\boldsymbol{r},t)\equiv\alpha\left(\boldsymbol{n}_{\mathcal{D}}\cdot\boldsymbol{E}_{\mathcal{D}}(\boldsymbol{r},t)\right), \tag{13}$$

where the electric field $\boldsymbol{E}_{\mathcal{D}}(\boldsymbol{r},t)$ is found reaching the surface $\Sigma$ from the dielectric interior and $\boldsymbol{n}_{\mathcal{D}}$ is the outward normal to the dielectric surface;

- a polarization current

$$\boldsymbol{J}_p(\boldsymbol{r},t)=\dot{\boldsymbol{P}}(\boldsymbol{r},t)=\alpha\dot{\boldsymbol{E}}(\boldsymbol{r},t). \tag{14}$$

The presence of surface charges implies a discontinuity of the normal component of the electric field around $\Sigma$, given by

$$\Delta E_n=\tilde{\sigma}_p. \tag{15}$$

Each of the candidate expressions for the momentum $\boldsymbol{p}^{(e.m.)}$ of an e.m. wave, considered in Section 1, implies a particular form for the corresponding resultant force $\boldsymbol{F}^{(e.m.)}$ exerted by the e.m. wave on the dielectric material, kept at rest, while it crosses it. This correspondence is given by Equation (10). In particular, from Equations (1) and (2) we get

$$\boldsymbol{F}_M^{(e.m.)}=\boldsymbol{F}_A^{(e.m.)}-\frac{\alpha}{c}\frac{\mathrm{d}}{\mathrm{d}t}\int_{\mathcal{D}}\boldsymbol{E}\times\boldsymbol{B}\,\mathrm{d}\boldsymbol{r}, \tag{16}$$

[3]For a discussion of the effects of nonlinearities see [12].

[4]The general case can be treated along the same lines without any trouble.

where $F_{M/A}^{(e.m.)}$ denotes the force exerted on $\mathcal{D}$ by the e.m. field, under the Minkowski or Abraham hypotheses respectively[5], while kept at rest.

In the next subsections I will discuss the forces associated with the Abraham or the Minkowski choice.

### 3.1. Forces Associated with the Abraham Momentum

In Ref. [8] it has been shown that the Abraham form of the momentum implies, for the force exerted by an e.m. wave interacting with an insulator, the exact expression

$$\boldsymbol{F}_A^{(e.m.)} = \frac{1}{2}\int_\Sigma \tilde{\sigma}_p\left(\boldsymbol{E}_\mathcal{V} + \boldsymbol{E}_\mathcal{D}\right)\mathrm{d}\Sigma + \frac{1}{c}\int_\mathcal{D} \boldsymbol{J}_p \times \boldsymbol{B}\,\mathrm{d}\boldsymbol{r} \tag{17}$$

$$= \frac{1}{2}\int_\Sigma \tilde{\sigma}_p\left(\boldsymbol{E}_\mathcal{V} + \boldsymbol{E}_\mathcal{D}\right)\mathrm{d}\Sigma + \frac{\alpha}{c}\int_\mathcal{D} \dot{\boldsymbol{E}} \times \boldsymbol{B}\,\mathrm{d}\boldsymbol{r}. \tag{18}$$

In Equations (17) and (18) $\boldsymbol{E}_\mathcal{V}$ and $\boldsymbol{E}_\mathcal{D}$ are the electric fields reached on the dielectric surface $\Sigma$ as a limit from the vacuum or the dielectric side respectively and have different values as a consequence of the discontinuities existing [13] at the dielectric boundaries, Equation (15). Equation (17) shows that, in the Abraham case, the force exerted on an ideal dielectric by an e.m. wave is the resultant of those acting on polarization, volume plus surface, charges. In particular the correct treatment needed in the presence of discontinuity surfaces [14] is reproduced.

### 3.2. Forces Associated with the Minkowski Momentum

If we apply the same strategy to determine the force acting on a dielectric, under the hypothesis that the momentum has the Minkowski form, we get, from Equation (16),

$$\begin{aligned}\boldsymbol{F}_M^{(e.m.)} &= \boldsymbol{F}_A^{(e.m.)} - \frac{\alpha}{c}\int_\mathcal{D} \dot{\boldsymbol{E}} \times \boldsymbol{B}\,\mathrm{d}\boldsymbol{r} - \frac{\alpha}{c}\int_\mathcal{D} \boldsymbol{E} \times \dot{\boldsymbol{B}}\,\mathrm{d}\boldsymbol{r} \\ &= \boldsymbol{F}_A^{(e.m.)} - \frac{1}{c}\int_\mathcal{D} \boldsymbol{J}_p \times \boldsymbol{B}\,\mathrm{d}\boldsymbol{r} + \alpha\int_\mathcal{D} \boldsymbol{E} \times \left(\nabla \times \boldsymbol{E}\right)\mathrm{d}\boldsymbol{r}.\end{aligned} \tag{19}$$

The last term in Equation (19) can be transformed as

$$\begin{aligned}\alpha\int_\mathcal{D} \boldsymbol{E} \times \left(\nabla \times \boldsymbol{E}\right)\mathrm{d}\boldsymbol{r} &= \frac{\alpha}{2}\int_\mathcal{D} \nabla\left(\boldsymbol{E}^2\right)\mathrm{d}\boldsymbol{r} - \alpha\int_\mathcal{D}\left(\boldsymbol{E}\cdot\nabla\right)\boldsymbol{E}\,\mathrm{d}\boldsymbol{r} \\ &= \frac{\alpha}{2}\int_\Sigma \boldsymbol{E}_\mathcal{D}^2 \boldsymbol{n}_\mathcal{D}\mathrm{d}\Sigma - \alpha\int_\Sigma\left(\boldsymbol{n}_\mathcal{D}\cdot\boldsymbol{E}_\mathcal{D}\right)\boldsymbol{E}_\mathcal{D}\mathrm{d}\Sigma,\end{aligned} \tag{20}$$

in the absence of free charges inside the uniform insulator.

Putting together Equations (17), (19) and (20) and taking into account the discontinuity of the electric field at the dielectric surface, Equation (15), we get, for the force exerted in the Minkowski case,

$$\boldsymbol{F}_M^{(e.m.)} = \frac{\alpha}{2}\int_\Sigma \boldsymbol{E}_\mathcal{D}^2 \boldsymbol{n}_\mathcal{D}\,\mathrm{d}\Sigma + \frac{1}{2}\int_\Sigma \tilde{\sigma}_p^2 \boldsymbol{n}_\mathcal{D}\,\mathrm{d}\Sigma. \tag{21}$$

Equation (21) shows that the Minkowski force is the resultant of forces which are locally orthogonal to the insulator surface.

## 4. The $\alpha$-Expansion of the Forces

In the presence of polarizability analytical computations are, in general, not possible. We will therefore resort to a systematic small $\alpha$-expansion scheme described in [8] [10].

Taking into account that $\tilde{\sigma}_p$ and the discontinuity of the electric field are already of order $\alpha$, we easily check that, up to order $\alpha$, the Abraham force, Equation (18), reduces to [8]

$$\boldsymbol{F}_A^{(e.m.)} \approx \alpha\int_\Sigma\left(\boldsymbol{E}_0\cdot\boldsymbol{n}_\mathcal{D}\right)\boldsymbol{E}_0\mathrm{d}\Sigma + \frac{\alpha}{c}\int_\mathcal{D} \dot{\boldsymbol{E}}_0 \times \boldsymbol{B}_0\mathrm{d}\boldsymbol{r} \tag{22}$$

[5]In this section I do not consider magnetic materials, so that I will not distinguish between $\boldsymbol{B}$ and $\boldsymbol{H}$. I will discuss the extension of these considerations to the magnetic case in Section 6.

$$= \frac{\alpha}{c}\frac{\mathrm{d}}{\mathrm{d}t}\int_{\mathcal{D}} \boldsymbol{E}_0 \times \boldsymbol{B}_0 \, \mathrm{d}\boldsymbol{r} + \frac{\alpha}{2}\int_{\Sigma} \boldsymbol{n}_{\mathcal{D}} \, \boldsymbol{E}_0^2 \, \mathrm{d}\Sigma, \tag{23}$$

where $\boldsymbol{E}_0$ and $\boldsymbol{B}_0$ are the unperturbed ($\alpha = 0$) electric and magnetic fields, propagating freely, without any influence from the dielectric material.

From Equations (23) and (16) we immediately get the corresponding expression of the force acting on the insulator under the Minkowski hypothesis, to order $\alpha$

$$\boldsymbol{F}_M^{(e.m.)} \approx \frac{\alpha}{2}\int_{\Sigma} \boldsymbol{n}_{\mathcal{D}} \boldsymbol{E}_0^2 \, \mathrm{d}\Sigma. \tag{24}$$

We are now ready to discuss the Snell's law within the Minkowski scheme.

## 5. The Snell's Law with the Minkowski Momentum

It has been shown in [8] that, up to order $\alpha$, the Snell's law is a consequence of Equation (23), which, at the same time, follows from the Abraham form of the e.m. momentum.

In a recent paper [15] the derivation given in [8] has been challenged and the, apparently contrasting, result was obtained that Snell's law is rather a consequence of the Minkowski form of the momentum, Equation (1). The analysis and resolution of this discrepancy is, in my opinion, rather instructive and shows that the validity of the Snell's law does not discriminate between the Abraham or Minkowski proposals.

In this section I follow the strategy of Ref. [8], valid through order $\alpha$, using the Minkowski form of the force, Equation (24), which, by the way, exhibits the local orthogonality of the force with respect to the boundary, necessary for the validity of the argument presented in [15].

I consider the setup, relevant for the discussion of the Snell's law, in which an e.m. wave packet with an initial momentum $\boldsymbol{p}_\gamma^{(0)}$, hits a dielectric, occupying the half space $z \geq 0$, at an angle $\hat{i}$ with respect to the $z$-axis.

The total Minkowski momentum $\left(\boldsymbol{p}_{\mathcal{D}}\right)_M$ transferred to the insulator is obtained by an argument very similar to the one explained in [8], sect. (3.3), and amounts to

$$\left(\boldsymbol{p}_{\mathcal{D}}\right)_M = \int_{-\infty}^{+\infty} \left(\boldsymbol{F}_{\mathcal{D}}\right)_M \mathrm{d}t = -\hat{\boldsymbol{z}}\frac{\alpha}{2\cos\hat{i}} p_\gamma^{(0)}, \tag{25}$$

where Equation (24) has been used. In Equation (25) $\hat{\boldsymbol{z}} = -\hat{\boldsymbol{n}}_{\mathcal{D}}$ denotes the unit vector along the positive z-axis. By momentum conservation and the absence of reflection[6], we are led to attribute to the e.m. wave, once inside the insulator, a momentum $\boldsymbol{p}'_\gamma$ such that

$$\boldsymbol{p}_\gamma^{(0)} = \boldsymbol{p}'_\gamma + \left(\boldsymbol{p}_{\mathcal{D}}\right)_M = \boldsymbol{p}'_\gamma - \hat{\boldsymbol{z}}\frac{\alpha}{2\cos\hat{i}} p_\gamma^{(0)}. \tag{26}$$

Equation (26) implies

$$\left(\boldsymbol{p}'_\gamma\right)^2 = (1+\alpha)\left(\boldsymbol{p}_\gamma^{(0)}\right)^2 = n^2\left(\boldsymbol{p}_\gamma^{(0)}\right)^2, \tag{27}$$

where we used, according to the $\alpha$-expansion,

$$n = \sqrt{\epsilon} \approx 1 + \frac{\alpha}{2}. \tag{28}$$

Consistently, Equation (27) reproduces the Minkowski expression for the momentum from which we started.

Moreover Equation (26), together with Equation (28), also gives

$$\sin\hat{i}' = \frac{\left(\boldsymbol{p}'_\gamma\right)_\perp}{p'_\gamma} = \frac{\left(\boldsymbol{p}_\gamma^{(0)}\right)_\perp}{\left(1+\frac{\alpha}{2}\right)\left(p_\gamma^{(0)}\right)} \approx \frac{1}{n}\frac{\left(\boldsymbol{p}_\gamma^{(0)}\right)_\perp}{\left(p_\gamma^{(0)}\right)} = \frac{1}{n}\sin\hat{i}, \tag{29}$$

where $\hat{i}'$ is the refraction angle. Equation (29) reproduces the Snell's law, up to order $\alpha$, starting from the conservation of the Minkowski momentum.

---

[6] I remind the reader that the computation is performed up to order $\alpha$.

The conclusion of this computation is that both the Abraham and the Minkowski forms of the momentum are compatible with the Snell's law. Of course there is a difference, which can be experimentally detected, between the forces acting on the insulator, while crossed by the e.m. wave, in the Abraham or the Minkowski case. In particular, while in the Minkowski case the force exerted by an e.m. wave, during crossing the dielectric boundary, is orthogonal to the boundary itself, this is not true in the case of the Abraham force, Equation (23).

## 6. Magnetic Materials

In this section we extend the considerations of Section 4 of [8] to a non conducting magnetic material, $\mathcal{M}$, both in the Abraham and in the Minkowski case.

### 6.1. Surface Currents and Discontinuities

In $\mathcal{M}$ a magnetization field $\boldsymbol{M}$ is present, which corresponds to a bulk magnetization current density

$$\boldsymbol{j}_M = c\nabla\times\boldsymbol{M}. \tag{30}$$

The Ampère equation then becomes, in the absence of "free currents",

$$\nabla\times\boldsymbol{B} = \nabla\times\boldsymbol{M} + \frac{1}{c}\dot{\boldsymbol{E}}, \tag{31}$$

which allows the introduction of the magnetic field $\boldsymbol{H}$

$$\boldsymbol{H} = \boldsymbol{B} - \boldsymbol{M}. \tag{32}$$

In the linear regime

$$\boldsymbol{B} = \mu\boldsymbol{H} \tag{33}$$

and we have

$$\boldsymbol{M} = \frac{\mu-1}{\mu}\boldsymbol{B} \equiv \frac{\delta\mu}{\mu}\boldsymbol{B}, \tag{34}$$

which defines the magnetic susceptibility $\delta\mu$.

At the boundary surface, $\Sigma$, separating $\mathcal{M}$ from the vacuum $\mathcal{V}$, we have a discontinuity in $\boldsymbol{B}_{II}$, the projection of the magnetic induction on the tangent plane to $\Sigma$, given by

$$\left(\boldsymbol{B}_{\mathcal{M}} - \boldsymbol{B}_{\mathcal{V}}\right)_{II} = \boldsymbol{M}_{II}, \tag{35}$$

where $\boldsymbol{M}_{II}$ and $\boldsymbol{B}_{II}$ denote the projections of $\mathcal{M}$ and $\mathcal{B}$ on the tangent plane to $\Sigma$. The normal component, $B_n$, is continuous across $\Sigma$

$$\Delta B_n = 0. \tag{36}$$

Equation (35) is equivalent to the discontinuity due to a surface current [14]

$$\boldsymbol{j}_{\Sigma} = c\boldsymbol{M}\times\boldsymbol{n}_{\mathcal{M}} = c\boldsymbol{M}_{II}\times\boldsymbol{n}_{\mathcal{M}}. \tag{37}$$

### 6.2. The Magnetic Forces in the Minkowski Case

Following Ref. [8] we consider the candidate form for the e.m. momentum as

$$\boldsymbol{p}_M \equiv \frac{1}{c}\int_{\mathcal{R}_\infty\ominus\mathcal{L}_\delta}\boldsymbol{D}\times\boldsymbol{B}\mathrm{d}\boldsymbol{r}, \tag{38}$$

where, in order to deal with the dicontinuities, we exclude a thin region $\mathcal{L}_\delta$ of width $\delta$ around the surface of discontinuity $\Sigma$. In this way we can compute

$$\frac{\mathrm{d}\boldsymbol{p}_M}{\mathrm{d}t} = \frac{1}{c}\int_{\mathcal{R}_\infty\ominus\mathcal{L}_\delta}\dot{\boldsymbol{D}}\times\boldsymbol{B}\mathrm{d}\boldsymbol{r} + \frac{1}{c}\int_{\mathcal{R}_\infty\ominus\mathcal{L}_\delta}\boldsymbol{D}\times\dot{\boldsymbol{B}}\mathrm{d}\boldsymbol{r}. \tag{39}$$

We have

$$\frac{\mathrm{d}\boldsymbol{p}_M}{\mathrm{d}t} = -\frac{1}{c}\int_{\mathcal{M}} \boldsymbol{j}_M \times \boldsymbol{B}\,\mathrm{d}\boldsymbol{r} + \int_{\mathcal{R}_\infty \ominus \mathcal{L}_\delta} (\nabla \times \boldsymbol{B}) \times \boldsymbol{B}\,\mathrm{d}\boldsymbol{r} - \int_{\mathcal{R}_\infty \ominus \mathcal{L}_\delta} \boldsymbol{D} \times (\nabla \times \boldsymbol{E})\,\mathrm{d}\boldsymbol{r}$$
$$= -\frac{1}{c}\int_{\mathcal{M}} \boldsymbol{j}_M \times \boldsymbol{B}\,\mathrm{d}\boldsymbol{r} - \int_\Sigma \left[\frac{1}{2}\boldsymbol{n}_{\mathcal{M}} \Delta \boldsymbol{B}^2 - (\boldsymbol{n}_{\mathcal{M}} \cdot \boldsymbol{B}) \Delta \boldsymbol{B}\right] \mathrm{d}\Sigma, \tag{40}$$

where contributions from the surfaces at infinity have been neglected, since we always consider situations in which the e.m. field differs from zero only in a finite region of space.

We have

$$\frac{\mathrm{d}\boldsymbol{p}_M}{\mathrm{d}t} = -\frac{1}{c}\int_{\mathcal{M}} \boldsymbol{j}_M \times \boldsymbol{B}\,\mathrm{d}\boldsymbol{r} + \int_\Sigma \boldsymbol{B}_n \boldsymbol{M}_{II}\,\mathrm{d}\Sigma - \int_\Sigma \boldsymbol{n}_{\mathcal{M}} \left[\frac{(\boldsymbol{B}_{\mathcal{M}} + \boldsymbol{B}_{\mathcal{V}})}{2} \cdot \boldsymbol{M}_{II}\right] \mathrm{d}\Sigma, \tag{41}$$

where Equations (35) and (36) have been used.

Equation (41) shows that the total force $\boldsymbol{F}_M^{(e.m.)}$, exerted by an e.m. wave on $\mathcal{M}$ under the hypothesis of the Minkowski momentum, is given, according to the scheme of Equation (10), by

$$\boldsymbol{F}_M^{(e.m.)} = \frac{1}{c}\int_{\mathcal{M}} \boldsymbol{j}_M \times \boldsymbol{B}\,\mathrm{d}\boldsymbol{r} - \int_\Sigma \boldsymbol{B}_n \boldsymbol{M}_{II}\,\mathrm{d}\Sigma + \int_\Sigma \boldsymbol{n}_{\mathcal{M}} \left[\frac{(\boldsymbol{B}_{\mathcal{M}} + \boldsymbol{B}_{\mathcal{V}})}{2} \cdot \boldsymbol{M}_{II}\right] \mathrm{d}\Sigma. \tag{42}$$

In situations of weak magnetization,

$$\mu - 1 \equiv \delta\mu \approx 0 \tag{43}$$

in which $(\delta\mu)^2$ is negligible, the discontinuity of $\boldsymbol{B}_{II}$ can be neglected in Equation (42), which therefore reduces to

$$\boldsymbol{F}_M^{(e.m.)} \approx \frac{1}{c}\int_{\mathcal{M}} \boldsymbol{j}_M \times \boldsymbol{B}\,\mathrm{d}\boldsymbol{r} + \frac{1}{c}\int_\Sigma \boldsymbol{j}_\Sigma \times \boldsymbol{B}\,\mathrm{d}\Sigma. \tag{44}$$

Equation (44) shows that the Minkowski choice, Equation (38), for the momentum corresponds to a force on magnetic materials simply described as the Lorentz force acting on surface and bulk magnetization currents.

### 6.3. The Magnetic Forces in the Abraham Case

We now consider, again for the same magnetic material $\mathcal{M}$, the Abraham form for the momentum

$$\boldsymbol{p}_A = \frac{1}{c}\int_{\mathcal{R}_\infty \ominus \mathcal{L}_\delta} \boldsymbol{E} \times \boldsymbol{H}\,\mathrm{d}\boldsymbol{r}, \tag{45}$$

again regularized by the omission of the $\mathcal{L}_\delta$ strip. The computation of the forces on $\mathcal{M}$, according to the scheme of Equation (10) gives, in this case,

$$\frac{\mathrm{d}\boldsymbol{p}_A}{\mathrm{d}t} = \frac{1}{c}\int_{\mathcal{R}_\infty \ominus \mathcal{L}_\delta} \dot{\boldsymbol{E}} \times \boldsymbol{H}\,\mathrm{d}\boldsymbol{r} + \frac{1}{c}\int_{\mathcal{R}_\infty \ominus \mathcal{L}_\delta} \boldsymbol{E} \times \dot{\boldsymbol{H}}\,\mathrm{d}\boldsymbol{r}$$
$$= \int_{\mathcal{R}_\infty \ominus \mathcal{L}_\delta} (\nabla \times \boldsymbol{H}) \times \boldsymbol{H}\,\mathrm{d}\boldsymbol{r} - \frac{1}{c}\int_{\mathcal{M}} \boldsymbol{E} \times \dot{\boldsymbol{M}}\,\mathrm{d}\boldsymbol{r}, \tag{46}$$

where we used Equation (32) and we have again neglected contributions from the surfaces at infinity.

The first term in Equation (46) only requires the knowledge of the discontinuity conditions

$$(H_{\mathcal{M}})_n - (H_{\mathcal{V}})_n = -M_n \tag{47}$$

$$\Delta \boldsymbol{H}_{II} = 0. \tag{48}$$

We have

$$\begin{aligned}
&\int_{\mathcal{R}_\infty \ominus \mathcal{L}_\delta} (\nabla \times \boldsymbol{H}) \times \boldsymbol{H}\, \mathrm{d}\boldsymbol{r} \\
&= \int_\Sigma \mathrm{d}\Sigma \left[ (\boldsymbol{n}_{\mathcal{M}} \cdot \boldsymbol{H}_{\mathcal{M}}) \boldsymbol{H}_M - \frac{1}{2} n_{\mathcal{M}} (\boldsymbol{H}_{\mathcal{M}}^2) + (\boldsymbol{n}_{\mathcal{V}} \cdot \boldsymbol{H}_{\mathcal{V}}) \boldsymbol{H}_{\mathcal{V}} - \frac{1}{2} \boldsymbol{n}_{\mathcal{V}} (\boldsymbol{H}_{\mathcal{V}}^2) \right] \\
&= \int_\Sigma \mathrm{d}\Sigma \left[ H_{\mathcal{M}n} \boldsymbol{H}_{\mathcal{M}} - H_{\mathcal{V}n} \boldsymbol{H}_{\mathcal{V}} + \frac{1}{2} \boldsymbol{n}_{\mathcal{M}} (\boldsymbol{H}_{\mathcal{V}}^2 - \boldsymbol{H}_M^2) \right] \\
&= \int_\Sigma \mathrm{d}\Sigma \left[ -M_n \boldsymbol{H}_{\mathcal{M}II} - \boldsymbol{n}_{\mathcal{M}} (H_{\mathcal{M}n} + H_{\mathcal{V}n}) M_n + \frac{1}{2} \boldsymbol{n}_{\mathcal{M}} (H_{\mathcal{M}n} + H_{\mathcal{V}n}) M_n \right] \\
&= \int_\Sigma \mathrm{d}\Sigma \left[ -M_n \boldsymbol{H}_{\mathcal{M}II} - \frac{1}{2} \boldsymbol{n}_{\mathcal{M}} (H_{\mathcal{M}n} + H_{\mathcal{V}n}) M_n \right],
\end{aligned} \tag{49}$$

and the expression of the force, in the Abraham case, is

$$\boldsymbol{F}_A^{(e.m.)} = \frac{1}{c} \int_{\mathcal{M}} \boldsymbol{E} \times \dot{\boldsymbol{M}} \mathrm{d}\boldsymbol{r} + \int_\Sigma \mathrm{d}\Sigma \left[ M_n \boldsymbol{H}_{M\,II} + \frac{1}{2} \boldsymbol{n}_M (H_{\mathcal{M}n} + H_{\mathcal{V}n}) M_n \right]. \tag{50}$$

Equation (50) has a simple interpretation which can be clarified considering the case of a very small $\delta\mu$. In fact in such a limit the discontinuity of $H_n$ across $\Sigma$ can be neglected and Equation (50) simplifies to

$$\begin{aligned}
\boldsymbol{F}_A^{(e.m.)} &\approx \frac{1}{c} \int_{\mathcal{M}} \boldsymbol{E} \times \dot{\boldsymbol{M}} \mathrm{d}\boldsymbol{r} + \int_\Sigma \mathrm{d}\Sigma \left[ M_n \boldsymbol{H}_{II} + \boldsymbol{n}_{\mathcal{M}} M_n H_n \right] \\
&= \frac{1}{c} \int_{\mathcal{M}} \boldsymbol{E} \times \dot{\boldsymbol{M}} \mathrm{d}\boldsymbol{r} + \int_\Sigma \mathrm{d}\Sigma M_n \boldsymbol{H}.
\end{aligned} \tag{51}$$

Equation (51) represents the interaction of the electric field with a "magnetic" current $\dot{\boldsymbol{M}}$, via a dual Lorentz force and the interaction of the magnetic field $\boldsymbol{H}$ with a distribution of magnetic charges, with a surface density $M_n$. It is interesting to notice that this expression for the force on magnetic materials is identical with the one proposed in Ref. [16].

## 7. Conclusions

In this paper I examined some questions related to the Abraham-Minkowski controversy about the momentum to be attributed to an e.m. wave packet inside a polarizable material. The analysis has been performed from the point of view of the forces experienced by the (dielectric or magnetic) material and is based on a careful treatment of the surface charges and currents and the consequent discontinuities of the electric and magnetic fields. While these results do not require any approximation, except for the idealized definition of the involved materials, we also consider the expansion in powers of the susceptibility $\alpha$, which allows to show that, at least up to first order in $\alpha$, the validity of the Snell's law for the refraction of e.m. waves is not able to discriminate between the Abraham and the Minkowski proposals.

The resulting picture is that the Abraham momentum implies that the action of an e.m. field on a dielectric can be understood quite naturally as the resultant of the usual e.m. forces acting on the polarization charges and currents. In the Minkowski case the form of the force on a dielectric material is not as natural. For magnetic materials, the Abraham form of the momentum implies a picture of the force as due to the action of the magnetic field $\boldsymbol{H}$ on a distribution of magnetic charges determined by the magnetization, along the lines discussed in Ref. [16]. The Minkowski form of the momentum, on the contrary, gives an expression of the force which can be interpreted as the Lorentz force exerted by the magnetic induction $\boldsymbol{B}$ on the bulk and the surface polarization currents.

The choice between these alternatives can only come from experiments.